\documentclass[11pt]{article}
\usepackage{hyperref}

\begin{document}


\begin{center}
{\Large \bf Kaluza-Klein Reduction of a Quadratic Curvature Model} 

\vspace{7mm}

S. Ba{\c s}kal
\footnote{electronic address:
baskal@newton.physics.metu.edu.tr} 
and H. Kuyrukcu
\footnote{electronic address: halil@metu.edu.tr}

\vspace{3mm}

Physics Department, Middle East Technical University\\
06531, Ankara, Turkey
\end{center}

\vspace{6mm}

\begin{abstract}
Palatini variational principle is implemented on a five dimensional
quadratic curvature gravity model, rendering two sets of equations which can be
interpreted as the field equations and the stress-energy tensor.  Unification 
of gravity with electromagnetism and the scalar dilaton field is achieved 
through the Kaluza-Klein dimensional reduction mechanism.  The reduced curvature 
invariant, field equations and the stress-energy tensor in four dimensional 
spacetime are obtained.  The structure of the interactions among the constituent 
fields is exhibited in detail.  It is shown that the Lorentz force naturally 
emerges from the reduced field equations and the equations of the standard 
Kaluza-Klein theory is demonstrated to be intrinsically contained in this model.
\end{abstract}

\section{Introduction}
Theories in dimensions higher than four seem to be promising candidates for
the ultimate unification of fundamental forces. 
The almost ninety years old Kaluza-Klein 
(KK) theory unifying electromagnetism with gravity in 5D \cite{kaluza21}, 
currently appears as a part of more involved models \cite{duff94,canfora10}, 
may still have some aspects that deserve to be investigated in their own right.  
As is well known, the standard KK (SKK) theory is obtained by the metric 
variations of the 5D Einstein-Hilbert (EH) action \cite{klein26}. 
Since then there have been many alternative approaches to the theory  
dealing with various types of actions \cite{allah10}, including those that 
contain dimensionally extended Euler 
densities \cite{muller86,muller88,huang88,dereli90,soleng95}.  
Usually, the field equations are obtained by metric variations of the action.
In this work, we shall deviate from this common practice and  
employ the Palatini variational principle which takes variations both with
respect to the connection and to the metric by treating them as 
independent variables \cite{sotiriou10}.   

If the requirement on the order of the derivatives of the metric in the
equations is released, then we can proceed with the simplest gravity model
and consider an action which is quadratic in the curvature.
Motivations for choosing such an action is more substantial than seeking 
for simplicity or pursuing for analogy with other gauge theories whose 
field equations and the stress-energy (SE) tensors are derived from an 
action quadratic in their fields.
In addition to its natural emergence as the leading term in string 
generated gravity models in their low energy limit \cite{deser85}, 
when coupled to matter its renormalizability problems become much 
less severe \cite{stelle77}.    
 
Here, we shall investigate the consequences of the KK reduction
mechanism on the 5D field equations and the SE tensor 
obtained from implementing the Palatini variational principle to a 
quadratic curvature 5D gravity model.

\section{The quadratic curvature model in five dimensions}
The basic operating mechanism of the KK theory can safely be viewed as
a spontaneous compactification of the five dimensional spacetime $M_{5}$
(with coordinates $(x^{a},y)$) to $M_{4}\times S^{1}$ while the 5D
Poincar{\' e}
symmetry $P_{4}$ of $M_{5}$ is spontaneously broken to  $P_{5}\times U(1)$.
Here, $M_{4}$ is the actual spacetime and $S^{1}$ has the topology of
a circle whose radius is assumed to be at the order of the Plank length.
The line element on $M_{5}$ is written as
\begin{equation}
ds_{5}^{2}=\hat g_{AB} \, \hat e^{A}\otimes \hat e^{B},
\end{equation}
with its signature adopted as (-,+,+,+,+).
The capital indices $A,B,...$ assume the values $0, 1, 2, 3, 5$ 
and the lower case indices $j,k, ...$ run from $0$ to $3$.
We shall be working in the horizontal lift basis (HLB) which will prove to
be convenient for our purposes.  An adequate amount of detail 
in the context of the SKK theory coupled to 
the Dirac field can be found in \cite{macias91}. 
In a more general framework it is referred
as an anholonomic basis and elucidated in \cite{misner73}.

Then the metric $\hat g_{AB}$ takes the form
\begin{equation}
\left(
\begin{array}{cc}
g_{ij}  & 0 \\
0 & 1
\end{array}
\right)
\end{equation}
with the basis
\begin{equation}
\hat e^{j}(x^{a},y)=dx^{j}, \qquad
\hat e^{5}(x^{a},y)=\varphi \, (x^{a})(dy+A_{k}(x^{a})dx^{k}).
\end{equation}

The SKK theory uses the 5D EH action
\begin{equation}
\hat S_{EH}=\int  (-\hat g)^{\frac{1}{2}}\, \hat{R}\, d^{5}x
\end{equation}
where $\hat{R}$ is the 5D curvature scalar.  The field equations
of the theory $\hat{R}_{AB}=0$, are obtained through metric variations. 
Using the KK reduction mechanism they 
are expressed as \cite{wesson99}:
\begin{equation}\label{skk}
\begin{array}{l}
K1_{ab}\equiv
R_{ab}-\frac{1}{2}\varphi^{2}F_{ak}F_{b}\,^{k}+\varphi^{-1}D_{a}\varphi_{b}=0\\
K2_{k} \equiv D_{j}F^{j}\,_{k}+3\varphi^{-1}\varphi^{j}F_{jk}=0\\
K3 \equiv D_{j}\varphi^{j}-\frac{1}{4}\varphi^{3}F_{jk}F^{jk}=0.
\end{array}
\end{equation}
Here, $R_{ab}$ is the 4D Ricci tensor,
$F_{ab}=\partial_{a}A_{b}-\partial_{b}A_{a}$ is the electromagnetic (EM) field
strength tensor
and $\varphi$ is the dilaton field.  For convenience, we also give the
reduced forms of the 5D curvature tensor in HLB:
\begin{equation}\label{riem}
\begin{array}{l}
\hat{R}_{abmn}=R_{abmn}-\frac{1}{4}\varphi^{2}
(2 F_{ab}F_{mn}+F_{am}F_{bn}-F_{an}F_{bm}) \\[1ex]
\hat{R}_{5bmn}=\frac{1}{2}\varphi D_{b}F_{mn}
+\frac{1}{2}( 2\varphi_{b} F_{mn}
+\varphi_{m} F_{bn}-\varphi_{n}F_{bm})\\ [1ex]
\hat{R}_{a5m5}=-\varphi^{-1}D_{m}\varphi_{a}
-\frac{1}{4}\varphi^{2}F_{aj}F^{j}\,_{m}.
\end{array}
\end{equation}

In this article, we shall consider an alternative 5D action  
\begin{equation}\label{action}
\hat S=\int  {\mathcal L} \, d^{5}x
\end{equation}
where
\begin{equation}
{\mathcal L}=  (-\hat g)^{\frac{1}{2}} \, \hat{R}^{JKMN}\hat{R}_{JKMN}
\end{equation}
which is quadratic in the 5D curvature tensor.  To reduce the quadratic
invariant $\hat \mathcal{I}_{q}$
into the actual 4D spacetime, it is expanded as 
\begin{equation}\label{expr2}
\begin{array}{ll}
\hat \mathcal{I}_{q}=\hat{R}^{JKMN}\hat{R}_{JKMN}=
 \hat{R}^{jkmn}\hat{R}_{jkmn}
+4\hat{R}^{jkm5}\hat{R}_{jkm5}+4\hat{R}^{k5m5}\hat{R}_{k5m5}
\end{array}
\end{equation}
and the substitution of (\ref{riem}) into above is sufficient.
    
We adopt the Palatini approach, where the metric and the connection
variations are considered to be independent, thus producing two sets of
equations.  

Variations of the action (\ref{action}) with respect to the 
connection $\delta {\mathcal L} / \delta \Gamma^{A}\,_{BC}=0 $ renders 
\begin{eqnarray}\label{5fe}
\hat{D}_{K}\hat{R}^{K}\,_{B  MN}=0
\end{eqnarray}
and we interpret it as the field equations of the model.  
Varying with respect to the 5D metric 
\begin{equation}
\frac{(-\hat g)^{-1/2}}{2}
\frac{\delta {\mathcal L}}{\delta \hat g_{AB}}
            \equiv \hat{T}_{AB}
\end{equation}
gives
\begin{equation}\label{5set}
\hat{T}_{AB}=\hat{R}_{AKMN}\hat{R}_{B}\,^{KMN}
-\frac{1}{4}\hat{g}_{AB}\hat{R}_{JKMN}\hat{R}^{JKMN}
\end{equation}
and interpreted as the SE tensor of the model
which is symmetric and due to the field equations (\ref{5fe}) is 
covariantly conserved
\begin{equation}
\hat D_{K}\hat{T}^{K}\,_{B}=0.
\end{equation}

This approach is quite similar to that of the gauge theories
whose the field equations are obtained by varying with respect to the gauge 
potentials, while the SE tensor is obtained through the metric variations of
the action.  The gauge structure of gravity with an appropriate choice of the
gauge group is well established 
\cite{yang74,mansouri76,gronwald96}, where the Riemann tensor and the
connection are behaving like the curvature and the gauge potential,
respectively. 
Therefore, the implementation of the Palatini method on the action
(\ref{action}) can be regarded as a natural extension of a familiar 
approach to gravity.

\section{The Kaluza-Klein Reduction}
The reduced form of the quadratic invariant in (\ref{expr2}) becomes  
\begin{equation}\label{rlag}
\begin{array}{l}
\hat \mathcal{I}_{q}=R^{jkmn}R_{jkmn}-\frac{3}{2}\varphi^{2} R_{jkmn}F^{jk}F^{mn}
+\frac{3}{8}\varphi^{4}F_{jk}F^{jk}F_{mn}F^{mn} \\[1ex]
+\frac{5}{8}\varphi^{4}F_{jk}F^{km}F_{mn}F^{nj}
+\varphi^{2}(D_{k}F_{mn})(D^{k}F^{mn}) \\[1ex]
+6 ( \varphi_{k}\varphi^{k} F_{mn}F^{mn}
+ \varphi_{m}\varphi_{n}F^{mk}F^{n}\,_{k}) \\[1ex]
+ 4 \varphi
(\varphi^{m} F^{kn} + \varphi_{k} F^{mn})(D_{k}F_{mn}) \\[1ex]
+4\varphi^{-2}D_{m}\varphi_{n}D^{m}\varphi^{n}
-2 \varphi D_{m}\varphi_{n} F^{mk}F^{n}\,_{k},
\end{array}
\end{equation}
which has also been earlier evaluated in \cite{dereli90} using differential
forms, and appears to be different from above due to our contingent use of
gauge and gravitational Bianchi identities:
\begin{equation} 
D_{[k}F_{mn]}=0, \qquad D_{[k}R_{mn]ij}=0. 
\end{equation}
In the sequel, we shall also be using 
\begin{equation}
(D_{m}D_{n}-D_{n}D_{m})F^{i}\,_{j}=R^{i}\,_{kmn}F^{k}\,_{j}
                                  -R^{k}\,_{jmn}F^{i}\,_{k}
\end{equation}
and 
\begin{eqnarray*}
2R_{jkmn}F^{jm}F^{kn}=R_{jkmn}F^{jk}F^{mn}
\end{eqnarray*}
whenever they happen to be useful for our purposes.
 
\subsection{The Reduction of the Field Equations}
The field equations (\ref{5fe}) comprises four equations to be reduced
\begin{eqnarray}\label{fe1234}
\hat{D}_{K}\hat{R}^{K}\,_{bmn}=0, \quad
\hat{D}_{K}\hat{R}^{K}\,_{5mn}=0,\quad
\hat{D}_{K}\hat{R}^{K}\,_{b5n}=0, \quad
\hat{D}_{K}\hat{R}^{K}\,_{5m5}=0.
\end{eqnarray}
The first of the above equations becomes
\begin{equation}\label{rfe1}
\begin{array}{l}
D_{k}R^{k}\,_{bmn}
+\varphi^{2}
\left\{
\frac{1}{4}\left[F_{bm}D_{k}F^{k}\,_{n}-F_{bn}D_{k}F^{k}\,_{m}\right]
\right. \\[2ex]  \left.
\frac{1}{2}\left[F_{nk}D_{m}F^{k}\,_{b}
-F_{mk}D_{n}F^{k}\,_{b}-D_{k}(F^{k}\,_{b}F_{mn})\right]
\right\} \\ [2ex]
+\varphi^{-1}\varphi_{k} R^{k}\,_{bmn}  \\[2ex]
+\varphi^{-2}
(\varphi_{m} D_{n}\varphi_{b}- \varphi_{n} D_{m} \varphi_{b}) \\ [2ex]
-\frac{3}{4}\varphi \varphi_{k}
(2 F^{k}\,_{b}F_{mn} + F^{k}\,_{m}F_{bn} - F^{k}\,_{n}F_{bm})  \\ [2ex]
+\varphi (\varphi_{m} F_{bk}F^{k}\,_{n} - \varphi_{n} F_{bk}F^{k}\,_{m}) =0
\end{array}
\end{equation}
and the second equation is reduced as
\begin{equation}\label{rfe2}
\begin{array}{l}
- \frac{1}{2}\varphi (F^{jk}R_{jkmn}+D_{k}D^{k}F_{mn})
+ \frac{1}{4}\varphi^{3}F^{jk}(F_{jk}F_{mn}+2F_{jm}F_{kn})  \\[2ex]
+ (F_{n}\,^{k}D_{m}\varphi_{k}- F_{m}\,^{k}D_{n}\varphi_{k} )
 \\[2ex]
+ \frac{1}{2}
(\varphi_{n} D_{k}F^{k}\,_{m}-\varphi_{m}D_{k}F^{k}\,_{n} ) \\[2ex]
- \frac{3}{2} \varphi^{k} D_{k}F_{mn}
- F_{mn} D_{k}\varphi^{k}= 0 .
\end{array}
\end{equation}
The third equation becomes
\begin{equation}\label{rfe3}
\begin{array}{l}
\frac{1}{2}\varphi(F^{jk}R_{kbjn}+D_{k}D_{n}F^{k}\,_{b})
+\frac{1}{8}\varphi^{3} F^{jk}(F_{jk}F_{bn}+2F_{jb}F_{kn})
\\[2ex]
-F_{b}\,^{k} D_{k}\varphi_{n}
-\frac{1}{2} F_{bn} D_{k} \varphi^{k}   \\[2ex]
+\frac{3}{2}\varphi^{-1}\varphi_{k}
(\varphi _{n}F^{k}\,_{b}+\varphi_{b}F^{k}\,_{n})
\\[2ex]
+\frac{3}{2}\varphi_{k}D_{n}F^{k}\,_{b}
+\varphi_{n} D_{k}F^{k}\,_{b}
+\frac{1}{2}\varphi_{b}D_{k}F^{k}\,_{n} = 0.
\end{array}
\end{equation}
Similarly, through the reduction recipe last equation takes the following form 
\begin{equation}\label{rfe4}
\begin{array}{l}
\frac{1}{4}\varphi^{2}\left[F_{mj}D_{k}F^{kj}+D_{m}(F^{jk}F_{jk})\right]
\\[2ex]
+\varphi^{-2}\varphi^{j}D_{m}\varphi_{j}
-\varphi^{-1}D_{j}D^{j}\varphi_{m}\\[2ex]
-\frac{5}{4}\varphi \varphi_{j}F^{jk}F_{km}
-\frac{3}{4}\varphi \varphi_{m}F^{jk}F_{kj} = 0 .
\end{array}
\end{equation}
Considering the special case $\varphi=constant$, the equation above
simplifies as:
\begin{equation}
F_{mj}D_{k}F^{kj}+D_{m}(F^{jk}F_{jk})=0.
\end{equation}
Introducing $F^{2}=F^{jk}F_{jk}$ and $J^{j}=D_{k}F^{kj}$,
and interpreting the latter as the current density, it can be expressed as
\begin{equation}
F_{mj}J^{j}= -D_{m}F^{2}
\end{equation}
where, it can safely be interpreted as the "Lorentz force", which is
derived from a scalar $F^{2}$.  Lorentz force within the context of the 5D
geodesic equation has been presented earlier in \cite{kerner00}.

We shall reorganize the field equations from (\ref{rfe1}) to (\ref{rfe4}) 
by freely using identities. Then (\ref{rfe1}) becomes
\begin{equation}\label{arfe1}
\begin{array}{l}
D_{n}(R_{bm}-\frac{1}{2}\varphi^{2}F_{bk}F_{m}\,^{k}-\varphi^{-1}D_{m}\varphi_{b})
\\[2ex]
-D_{m}(R_{bn}-\frac{1}{2}\varphi^{2}F_{bk}F_{n}\,^{k}-\varphi^{-1}D_{n}\varphi_{b})
\\[2ex]
+\frac{1}{4} [ F_{bn}(D_{k}F^{k}\,_{m}+3\varphi^{-1}\varphi_{k}F^{k}\,_{m})
- F_{bm}(D_{k}F^{k}\,_{n}+3 \varphi^{-1} \varphi_{k}F^{k}\,_{n})
\\[2ex]
  +2F_{mn}(D_{k}F^{k}\,_{b}+3 \varphi^{-1} \varphi_{k}F^{k}\,_{b})]=0
 \end{array}
\end{equation}
and equation (\ref{rfe2}) is organized as
\begin{equation}\label{arfe2}
\begin{array}{l}
D_{m}(D_{k}F^{k}\,_{n}+3\varphi^{-1}\varphi_{k}F^{k}\,_{n})
-D_{n}(D_{k}F^{k}\,_{m}+3\varphi^{-1}\varphi_{k}F^{k}\,_{m})\\ [2ex]
+F_{m}\,^{k}(R_{kn}-\frac{1}{2}\varphi^{2}F_{kj}F_{n}\,^{j}-\varphi^{-1}D_{k}\varphi_{n})
\\ [2ex]
-F_{n}\,^{k}(R_{km}-\frac{1}{2}\varphi^{2}F_{kj}F_{m}\,^{j}-\varphi^{-1}D_{k}\varphi_{m})\\[2ex]
+2\varphi^{-1}F_{mn}(D_{j}\varphi^{j}-\frac{1}{4}\varphi^{3}F_{jk}F^{jk}) =
0.
\end{array}
\end{equation}
The third equation (\ref{rfe3}) is similarly rearranged as
\begin{equation}\label{arfe3}
\begin{array}{l}
\varphi D_{n}(D_{k}F^{k}\,_{b}+3\varphi^{-1}\varphi_{k}F^{k}\,_{b}) \\[2ex]
+\varphi_{b}(D_{k}F^{k}\,_{n}+3\varphi^{-1}\varphi_{k}F^{k}\,_{n})
+2 \varphi_{n}(D_{k}F^{k}\,_{b}+3\varphi^{-1}\varphi_{k}F^{k}\,_{b}) \\[2ex]
+F_{nb}(D_{k}\varphi^{k}-\frac{1}{4}\varphi^{3}F_{jk}F^{jk}) \\[2ex]
-\varphi F_{b}\,^{k}
(R_{kn}-\frac{1}{2}\varphi^{2}F_{kj}F_{n}\,^{j}-\varphi^{-1}D_{k}\varphi_{n})
= 0
\end{array}
\end{equation}
and the last equation (\ref{rfe4}) takes the form
\begin{equation}\label{arfe4}
\begin{array}{l}
D_{m}(D_{k}\varphi^{k}-\frac{1}{4}\varphi^{3}F_{jk}F^{jk})
\\[2ex]
-\frac{1}{4}\varphi^{3}F_{m}\,^{j}(D_{k}F^{k}\,_{j}+3\varphi^{-1}\varphi_{k}F^{k}\,_{j})
\\[2ex]
+\varphi_{j}(R_{jm}-\frac{1}{2}\varphi^{2}F_{jk}F_{m}\,^{k}-\varphi^{-1}D_{j}\varphi_{m})
= 0 .
\end{array}
\end{equation}
Looking closely into the above four equations, it is now possible to
recognize
the patterns of the SKK equations in (\ref{skk}).  
Equation (\ref{arfe1}) contains
the covariant derivative of $K1_{ab}$ and particular couplings of 
$K2_{a}$ and $K3$ with the EM field strength tensor.  Similarly,  
(\ref{arfe2}) and (\ref{arfe3}) govern $F_{mn}$ containing the covariant
derivative of $K2_{a}$ and have couplings of $F_{mn}$ and $\varphi$ with  
the whole $K$-set of (\ref{skk}).  The last equation (\ref{arfe4}), 
includes the covariant derivative of $K3$, basically
governs the dilaton field and has couplings of $K2_{a}$ and $K1_{ab}$ 
with $F_{mn}$ and $\varphi$.  It can be seen that neither $R_{jkmn}$ 
nor $R_{ab}$ has any couplings with the $K$-set.
Therefore, any solution to the SKK equations also solves the field equations 
(\ref{arfe1})-(\ref{arfe4}).

\subsection{The Reduction of the SE-Tensor}
The $\hat T_{ab}$ component of the reduced stress-energy tensor in 
(\ref{5set}) can be separated into
its trace-free and a non-vanishing trace part.  Its trace-free part is:
\begin{eqnarray}\label{tfse1}
\hat{T}_{ab}^{(tracefree)}&=&
R_{akmn}
R_{b}\,^{kmn}
-\frac{1}{4}g_{ab}R_{jkmn}R^{jkmn}
\nonumber \\
& &-\frac{3}{2}\varphi^{2}
F^{mn}\left\{ \frac{1}{2}(F_{a}\,^{k}R_{bkmn}+F_{b}\,^{k}R_{akmn})
-\frac{1}{4}g_{ab}F^{jk}R_{jkmn} \right\}
\nonumber \\
& &+\frac{3}{8}\varphi^{4}F^{2}
\left(F_{ak}F_{b}\,^{k}-\frac{1}{4}g_{ab}F_{jk}F^{jk}\right)
\nonumber
\\& &
+\frac{5}{8}\varphi^{4}\biggl(F_{am}F^{mk}F_{kn}F^{n}\,_{b}
-\frac{1}{4}g_{ab}(F_{jk}F^{km}F_{mn}F^{nj})\biggr)
\nonumber \\& &
+\varphi^{2}\biggl\{(D_{a}F_{mn})(D_{b}F^{mn})
-\frac{1}{4}g_{ab}(D_{k}F_{mn})(D^{k}F^{mn})\biggr\}
\nonumber \\&
&+4\varphi^{-2}\biggl\{(D_{a}\varphi_{m})(D_{b}\varphi^{m})
-\frac{1}{4}g_{ab}(D_{n}\varphi_{m})(D^{n}\varphi^{m})\biggr\}
\nonumber
\\&
&+3 \varphi_{m}\varphi^{m}
\left(F_{ak}F_{b}\,^{k}-\frac{1}{4}g_{ab}F_{jk}F^{jk}\right)
\nonumber \\&
&+3F^{2}\biggl(\varphi_{a}\varphi_{b}
-\frac{1}{4}g_{ab}\varphi_{k}\varphi^{k}\biggr)
\\&&
+3\varphi^{j}\varphi^{k}\Bigl(F_{aj}F_{bk}
-\frac{1}{4}g_{ab}F_{nj}F^{n}\,_{k}\Bigr) \nonumber
\\&
&+3\left\{\varphi_{k}F^{kj}\Bigl[
\frac{1}{2}(\varphi_{a}F_{bj}+\varphi_{b}F_{aj})
-\frac{1}{4}g_{ab}\varphi_{m}F^{m}\,_{j}\Bigr]\right\}\nonumber
\\&
&-4\varphi\biggl\{F^{jk}
\Bigl[\frac{1}{2}(\varphi_{a}D_{j}F_{kb}+\varphi_{b}D_{j}F_{ka})
-\frac{1}{4}g_{ab}\varphi^{n}D_{j}F_{kn}\Bigr]
\biggr\}\nonumber
\\& &
+4\varphi\biggl\{\varphi^{k}\Bigl[\frac{1}{2}
\Bigl(F_{a}\,^{n}D_{k}F_{bn}+F_{b}\,^{n}D_{k}F_{an}\Bigr)
-\frac{1}{4}g_{ab}F^{jk}D_{m}F_{jk}\Bigr]\biggr\}\nonumber
\\& &
+2\varphi\biggl\{\frac{1}{2}\Bigl[(D_{a}\varphi_{m})F_{bj}F^{jm}
+(D_{b}\varphi_{m})F_{aj}F^{jm}\Bigr]
-\frac{1}{4}g_{ab}(D_{n}\varphi_{m})F^{nk}F_{k}\,^{m}\biggr\}\nonumber
\end{eqnarray}
and the part with a non-vanishing trace is:
\begin{eqnarray}\label{tse1}
\hat{T}_{ab}^{(ta)}&=&
-\frac{1}{8}\varphi^{4}F_{am}F^{mk}F_{kn}F^{n}\,_{b}
+\frac{1}{2}\varphi^{2}(D_{m}F_{ak})(D^{m}F_{b}\,^{k})
-\frac{3}{4}\varphi^{2}(D_{a}F_{mn})(D_{b}F^{mn})\nonumber
\\& &
-2\varphi^{-2}(D_{a}\varphi_{m})(D_{b}\varphi^{m})
-\frac{3}{2}F^{2}\varphi_{a}\varphi_{b}
-\frac{3}{2}\varphi_{j}\varphi_{k}F_{a}\,^{j}F_{b}\,^{k}\nonumber
\\& &
-\frac{1}{2}\varphi F^{jm}(F_{aj}D_{b}\varphi_{m}+F_{bj}D_{a}\varphi_{m})
+\frac{1}{2}\varphi(\varphi_{a}D_{m}F_{nb}+\varphi_{b}D^{m}F_{na})
\nonumber \\&&
-\frac{1}{2}\varphi \,\varphi^{k}
F_{ak}(D_{m}F_{b}\,^{k}+F_{b}\,^{k}D_{m}F_{ak}).
\end{eqnarray}
The reduced $\hat T_{a5}$ component of the SE tensor can be written as:
\begin{eqnarray}\label{se2}
\hat{T}_{a5} & = &
3\varphi^{-1}F^{m}\,_{a}\varphi^{k}D_{m}\varphi_{k}\nonumber \\
&&
+(D_{m}\varphi_{k})D^{m}F^{k}\,_{a}
+R_{akmn}(\varphi^{k}F^{mn}+\varphi^{m}F^{kn}) \nonumber \\
&&
+\frac{1}{2}\varphi R_{akmn}D^{k}F^{mn}
+\frac{3}{4}\varphi^{2}F^{2}\varphi_{k}F^{k}\,_{a}\nonumber \\
&&
+\frac{1}{4}\varphi^{3}
(3F_{a}\,^{k}F^{mn}D_{m}F_{nk}+F^{k}\,_{n}F^{nm}D_{m}F_{ka}).
\end{eqnarray}
And the last component $\hat T_{55}$ is reduced in the following manner:
\begin{eqnarray}\label{se3}
\hat{T}_{55} & = &
\varphi^{-2}(D_{m}\varphi_{n})D^{m}\varphi^{n}
-\frac{1}{4}R_{jkmn}R^{jkmn}
+\frac{1}{2}\varphi F^{m}\,_{n}F^{nk}D_{k}\varphi_{m} \nonumber
\\
&&
+\frac{3}{8}\varphi^{2}R_{jkmn}F^{jk}F^{mn}
-\frac{1}{32}\varphi^{4}(3F^{4}+F_{jk}F^{km}F_{mn}F^{nj}).
\end{eqnarray}
The trace of the $\hat T_{AB}$ can be written as
\begin{equation}
\hat T=\hat g^{AB}T_{AB}=\hat T^{a}\,_{a}+\hat T^{5}\,_{5}
=-\frac{1}{4}\hat \mathcal{I}_{q}.
\end{equation}
Therefore, the SE tensor is trace-free only when the invariant $\hat
\mathcal{I}_{q}$ vanishes.

\section{Conclusion}
As one would expect the dimensionally reduced    
4D field equations are more involved than those of the usual
Kilmister-Yang (KY) type of gravity, containing non-minimal couplings
of Ricci tensor and the field tensor of EM along with a scalar dilaton
field.  We have explicitly shown that our field equations are
some specific combinations of the equations of the SKK
theory.

It is found out that the set of field equations also contain the Lorentz force 
in addition to some particular couplings with Maxwell's equations.  
As is well known, the theory of EM is only complete when the Lorentz force 
is also taken into account.

The stress energy tensor contains particular non-minimal couplings 
with the well known SE tensors of the KY type of gravity, EM and the 
dilaton field. 
Although they appear to be more complicated than the standard SE tensors of 
those fields when considered separately, it seems to be the price we have to pay 
for unification and thus for mutual interactions.  We have also analyzed the trace 
of the SE tensor and its conservation properties and provided the conditions for 
trace-freedom.

It is known that the KY pure gravity equations contain
non-physical solutions.  In dimensions higher than four, the introduction of
the Gauss-Bonnet action seemed to provide some relief.  
In the context of the KK theory, couplings with the EM and the dilaton field 
requires more equations governing the fields that may alleviate the problem.  
A through 
analysis on this matter and on the existence of any further possible remedies is 
left to a later study.

Finally we conclude that this model turned out to be more complete by 
accommodating the Lorentz force and an intrinsic inclusion of the solutions 
of the SKK model\cite{kuyrukcu10}.

\end{document}